\documentclass[iop]{emulateapj}
\usepackage{verbatim}
\usepackage{rotating}
\usepackage{graphicx}

\begin{document}
\title{Velocity Anisotropy and Shape Bias in the Caustic Technique}
\author{Daniel Gifford and Christopher J. Miller}
\affil{Department of Astronomy, University of Michigan, 500 Church St. Ann Arbor, MI USA 48109}

\begin{abstract}
We use the Millennium Simulation to quantify the statistical accuracy and precision of the escape velocity technique for measuring cluster-sized halo masses at $z \sim 0.1$. We show that in 3D, one can measure nearly unbiased ($<$4\%) halo masses ($> 1.5\times10^{14}$M$_{\odot}$h$^{-1}$) with 10\%-15\% scatter. Line-of-sight projection effects increase the scatter to $\sim$25\%, where we include the known velocity anisotropies. The classical ``caustic'' technique incorporates a calibration factor which is determined from N-body simulations. We derive and test a new implementation which eliminates the need for calibration and utilizes only the observables: the galaxy velocities with respect to the cluster mean $v$, the projected positions $r_p$, an estimate of the Navarro-Frenk-White (NFW) density concentration and an estimate of the velocity anisotropies, $\beta$. We find that differences between the potential and density NFW concentrations induce a 10\% bias in the caustic masses. We also find that large (100\%) systematic errors in the observed ensemble average velocity anisotropies and concentrations translate to small (5\%-10\%) biases in the inferred masses.  
\end{abstract}

\section{Introduction}
\label{sec:introduction}

        Under Newtonian dynamics, the escape velocity is related to the gravitational potential of the system,
        \begin{equation}
            v_{esc}^2(r) = -2\Phi(r) .
        \label{eq:escape}
        \end{equation}
        If the dynamics of the system are controlled by the gravitational potential, tracers which cannot escape the potential exist in a well-defined region of radius/velocity ($r-v$) phase space. The extrema of the velocities in this phase space define a surface, the escape velocity profile,  $v_{esc}(r)$, which can be observed in projected sky coordinates. Given the observed $v_{esc}(r)$, this ``caustic'' technique allows one to infer the mass profile of a cluster to well beyond the virial radius \citep{Diaferio97}.

        With the latest deployments of wide-field ground-based multi-object spectrographs like VIMOS on the VLT \citep{LeFevre03}; IMACS on Magellan \citep{Dressler11}; HECTOSpec\footnote{http://www.cfa.harvard.edu/mmti/hectospec} on the MMT we are beginning to see large spectroscopic follow-up data sets of galaxy clusters. As a consequence, the caustic technique has become more widely adopted. 
        
        \citet{Geller13} compare caustic to weak lensing mass profiles and find agreement to within $30\%$ around a virial radius. \citet{Lemze09} perform a dynamical study of the cluster A1689 and find good agreement between the caustic mass profiles and both the weak lensing and X-ray inferred mass profiles. \citet{Rines12} measure the caustic mass profiles to large radii to estimate the ultimate halo mass in clusters, which includes all mass bound to halos in a future $\Lambda$CDM universe. \citet{Andreon10} utilize caustic masses to help calibrate the $M_{200}$-richness relation alongside mass estimates from velocity dispersion scaling relations. New deep imaging surveys like CLASH on the Hubble Space Telescope have been awarded a significant amount of Very Large Telescope (VLT) time to collect spectroscopy, in part to study the dynamical and caustic masses of clusters \citep{Postman12}. And of course there are a variety of planned large-scale spectroscopy programs both from the ground (BigBoss\footnote{http://bigboss.lbl.gov/}) and space (EUCLID\footnote{http://sci.esa.int/euclid}). These future efforts could enable caustic masses to be measured for many thousands of galaxy clusters.
      
        \citet{Gifford13a} \defcitealias{Gifford13a}{GMK} (hereafter GMK) used the Millennium Simulation \citep{Springel05} to show that cluster-sized caustic masses within a projected  $r_{200}$ (the radius which contains 200 times the critical density) are more precise and more accurate than virial masses measured from their projected velocity dispersions. However, the implementation of the escape velocity technique employs a number of steps which result in masses that are calibrated to the N-body simulation. Cluster masses based on the traditional caustic technique vary by 30\% depending on which calibration is used \citep{Diaferio97,Diaferio99,Serra11}. In this paper, we clarify where these calibrations are incorporated into the theory and we assess their validity and impact on the inferred masses. We also present a variation on the original escape velocity caustic technique which eliminates the calibration.

\section{Theory}
\label{sec:theory}

        Consider a mass distribution described by an NFW profile such that the mass density $\rho$ and the potential $\Phi$ radial profiles are:
        \begin{eqnarray}
        \rho(r) &=& \frac{\rho_0}{(r/r_0)(1+r/r_0)^2} \nonumber \\ 
        \Phi(r) &=& -\frac{4\pi G \rho_0 (r_0)^2 \ln(1+r/r_0)}{r/r_0}
        \label{eq:nfw_rhopot}
        \end{eqnarray}
        where $\rho_0$ is the normalization and $r_0$ is the NFW scale radius \citep{NFW97}. This is an example of a density - potential \emph{pair} which share the same values for the shape parameters $\rho_0$ and $r_0$ and are related via the Poisson equation, $\nabla^2\Phi(r) = 4 \pi G \rho (r)$. We can write the NFW-inferred spherical mass differential as: 
      \begin{eqnarray}
        \frac{dm}{dr} &=& 4\pi\rho (r) r^2  \nonumber \\
        G\frac{dm}{dr} &=& -\Phi(r) \left (\frac{(r/r_0)^2}{(1+r/r_0)^2 \ln(1+r/r_0)} \right )
        \label{eq:m_caus_diaferio}
        \end{eqnarray}
         where the unknowns are the gravitational potential $\Phi (r)$ and the scale $r_0$. This is a key step in our escape velocity technique, where we have equated the parameter $\rho_0$ in equations \ref{eq:nfw_rhopot}. The NFW parameter $\rho_0$ sets the absolute scale for the mass density. The other NFW parameter $r_0$ defines the scale radius and is observable in projected data, assuming light traces mass.
        
        We use equation \ref{eq:escape} to re-write equation \ref{eq:m_caus_diaferio} as:
        \begin{equation}
        G M(<R) = \int_0^R \hat{\mathcal{F}}(r) v^2_{esc}(r) dr 
        \label{eq:m_caustic_nfw}
        \end{equation}
        where 
        \begin{equation}
        \hat{\mathcal{F}}(r) =\frac{(r/r_0)^2}{(1+r/r_0)^2 \ln(1+r/r_0)} 
        \label{eq:F_r_nfw}
        \end{equation}
where the unknowns are the scale radius $r_0$ and the escape velocity $v^2_{esc}(r)$,  which is measured from the extrema in the radius and velocity ($r-v$) phase-space data. 

        More precisely, our estimate $\hat{\mathcal{F}}(r)$ should actually be:
        \begin{equation}
        \mathcal{F}(r) = -2 \pi G \frac{\rho (r) r^2}{\Phi(r)}
        \label{eq:F_r}
        \end{equation}
where $\rho(r)$ and $\Phi(r)$ are are the spherically averaged density and potential profiles (see \citet{Diaferio97} or \citetalias{Gifford13a}). The difference between $\mathcal{F}(r)$ and $\hat{\mathcal{F}}(r)$ is that the former uses an exact profile for the densities and the potentials, while the latter assumes that only the density profile can be measured and that the potential has the same NFW shape parameters (i.e., concentration and scale) as the potential. We discuss whether or not this NFW-shape assumption holds in Section \ref{sec:NFW_shapes}.

         In projected data, we measure the velocities along the line-of-sight (l.o.s.), and so
         \begin{equation}
        \langle v^2_{esc,los}\rangle (r) = \frac{(1-\beta(r))}{(3-2\beta(r))}\langle v_{esc}^2\rangle(r) = (g(\beta(r)))^{-1}\langle v_{esc}^2\rangle(r)
        \label{eq:v_proj}
        \end{equation}
         where the $\beta$ is the standard velocity anisotropy parameter. 
         
         In the classical implementation of the caustic technique,  the average $\mathcal{F}_{\beta} = \langle g(\beta(r)) \mathcal{F}(r) \rangle$ is measured within N-body simulations, and then applied to real data \citep{Rines12,Geller13}.  In the literature, $0.5 < \mathcal{F}_{\beta} < 0.7$ \citep{Diaferio97,Diaferio99,Serra11,Gifford13a}. Since $\mathcal{F}_{\beta}$ enters into the equation as being directly proportional to the mass, we must know it to high accuracy if escape velocity masses are to be used in cosmological analyses.

        A goal of this paper is to drop the requirement that $\mathcal{F}_{\beta}$ be calibrated from simulations. We assume that clusters are NFW density-potential {\it pairs} and apply equation \ref{eq:m_caustic_nfw} directly. The unknowns, $r_0$ and $\langle v^2_{esc,los} \rangle$ and $\beta$ are constrained from observed data \citep{LinMohr04,Carlberg97,Wojtak10,Budzynski12,Diaferio97,Geller13,Rines12,Lemze09,Biviano09,Host09}.

        A final calibration in standard escape-velocity technique is that of the iso-density surface in $r-v$ space which defines the average escape velocity, $\langle v_{esc}^2\rangle (r)$. The density-weighted average escape velocity inside radius $R$ is:
        \begin{equation}
        \langle v^2_{esc}(<R) \rangle = \frac{\int_{0}^{R} d^3{\bf x} \rho({\bf x}) v_{esc}^2({\bf x})}{\int_{0}^{R} d^3{\bf x} \rho({\bf x})} = -2 \frac{\int_{0}^{R} d^3{\bf x} \rho({\bf x}) \Phi ({\bf x})}{M(<R)} 
        \label{eq:avg_esc1}
        \end{equation}
        where we have used equation \ref{eq:escape}. The integral in the numerator on the right-hand side of equation \ref{eq:avg_esc1} is twice the total potential energy of the system or $2W$ \citep{Binney87}, which leads to:
        \begin{equation}
        \langle v^2_{esc} \rangle = -\frac{4 W(<R)}{M(<R)}
        \label{eq:avg_esc2}
        \end{equation}
        where $W$ and $M$ are the total potential energy and mass of the system within the radius $R$. 
        
        One often defines the following relation between the fraction of the total kinetic over the potential energy to that expected from a virialized halo:
        \begin{equation}
        b = 1 + \frac{2\textrm{T}}{\textrm{W}}
        \label{eq:bind_frac}
        \end{equation}
        where $T$ is the total kinetic energy. If we express the total kinetic energy of the system as $T = 1/2 M \langle v^{2} \rangle$ equation \ref{eq:avg_esc2} becomes:
        \begin{equation}
        \langle v^2_{esc} (<R) \rangle = -\frac{4 \langle v^2 (<R) \rangle}{b-1}
        \label{eq:vesc_sigma}
       \end{equation}
        where the average quantities are measured within the same radius, $R$. The standard calibration assumes that  $\langle b \rangle = 0$ in a virialized and isolated halo such that $2T = -W$. Thus $\langle v_{esc}^2 \rangle = 4 \langle v^{2} \rangle$, such that the escape velocity phase-space surface is calibrated through a measurement of the velocity dispersion.
              
       In this section, we have clarified where the calibration steps enter into the standard caustic analysis. The calibration includes the term $\mathcal{F}_{\beta}$, which is directly proportional to the estimate of the mass. This term comprises two parts: $\mathcal{F}(r)$ in Equation \ref{eq:F_r} and $g(\beta(r))$ in Equation \ref{eq:v_proj}. The other calibration step occurs from $\langle b \rangle$ in Equation \ref{eq:vesc_sigma}, which decides the iso-density contour in the $r-v$ phase-space that defines the escape velocity. We have also presented a derivation of the caustic technique which does not require a calibration of  $\mathcal{F}_{\beta}$, but which assumes an NFW density-potential pair and uses the observables in Equation \ref{eq:m_caustic_nfw}.
       
\section{Testing the Theory}
\label{sec:test}
        \begin{figure}
        \plotone{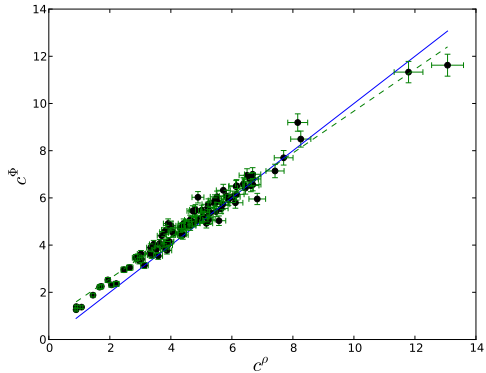}
        \caption{A comparison of the concentration measured via the density profile with $c^{\rho} = r_{200}/r_{0}^{\rho}$ and the concentration measured via the potential profile with $c^{\Phi} = r_{200}/r_0^{\Phi}$. The blue line is unity and the green dashed line is the fit to the relationship with a slope = 0.89 intercept = 0.80.}
        \label{fig:r_0_compare}
        \end{figure}        

We apply the caustic technique to 100 Millennium halos with masses $M_{200} > 1\times10^{14}$M$_{\odot}$h$^{-1}$ and $z < 0.1$ where $h = H_0$/100 km s$^{-1}$Mpc$^{-1}$. In 3D we use the particle positions and velocities. In the projected analyses we use the \citet{Guo11} semi-analytic galaxies within 30h$^{-1}$Mpc of the halo centers projected along a random line-of-sight.  These volumes are large enough to incorporate realistic projection effects. The limits on the projected phase-space velocities are $\pm 3000$km/s relative to the halo velocity centroids, whereas the typical escape velocities are $\sim 1500$km/s. 
        
\subsection{The NFW shapes}\label{sec:NFW_shapes}
        
    In order to drop the N-body calibration of $\mathcal{F}_{\beta}$, we assume that the NFW densities and potentials have the same parameters. This is expected if the matter distribution is concentric with the iso-potential surfaces (e.g., as in spherical symmetry; see also the classical potential solutions for homogeneous density distributions in \citet{Chandra69} and \citet{Binney87}). However, \citet{Conway00} provide exact closed-form Newtonian potential solutions to an infinite family of heterogeneous spheroids and find that the densities are generally not constant on the iso-potential bounding surfaces. In other words, while both the potential and density distribution could have the same general functional form like an NFW, they need not have identical shape parameters. 
    
    We compare the NFW density/potential shapes by first fitting the NFW density profile and determining  $\rho_0$ and $r_0$ for each halo. The gravitational potentials are measured exactly through summation of $\frac{Gm_i}{\bf r_i}$ and then fit with an NFW using $\rho_0$ measured from the density, but allowing the potential scale parameter $r_0$ to be a free parameter.
 
    In Figure \ref{fig:r_0_compare}, we compare the NFW concentrations $c^{\rho,\Phi} = r_{200}/r_0^{\rho,\Phi}$, where $r_{200}$ is the radius which contains a density corresponding to 200$\times$ the critical density. We find that the potentials have slightly higher concentrations than the densities. This difference suggests that our systems are not density-potential pairs which are simply related via spherical solutions to the Poisson equation. Because of this, we expect that using equation \ref{eq:m_caustic_nfw} will return an incorrect mass estimate due to its assumption of shape similarity in the density and potential profiles.
    
    \begin{figure*}
    \plotone{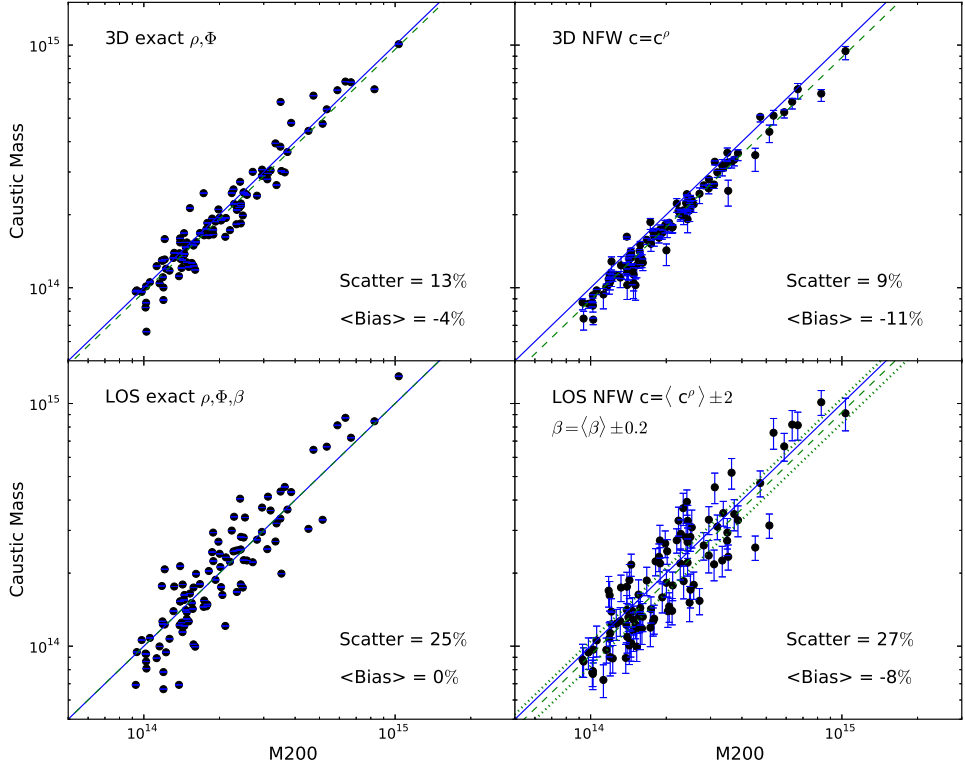}
    \caption{{\bf Top Left:} $M_{200}$ vs the 3D caustic mass estimated inside $r_{200}$ calculated using the exact potential and density profiles (see equation \ref{eq:F_r}). {\bf Top Right:} Caustic masses using NFW fits to the 3D density profiles (equation \ref{eq:m_caustic_nfw}).  The induced bias is expected from Figure \ref{fig:r_0_compare}. In all panels the solid blue line is unity and the green dashed line represents the average bias of the sample with slope unity. {\bf Bottom Left:} $M_{200}$ vs the line-of-sight caustic mass estimated inside a projected $r_{200}$. As in the top panel, we use the particle potential and density profiles, but now include the particle anisotropy profiles as well. The increased scatter is due to the line-of-sight projections which induce scatter into the velocity dispersions. {\bf Bottom Right:} Projected caustic masses based on an NFW density profile with a single sample concentration of $\langle c \rangle = 5 \pm 2$ and a single sample $\langle \beta \rangle = 0.2 \pm 0.2$. These large uncertainties do not add appreciably to the scatter induced by the line-of-sight projection effects. Mass biases induced by systematic errors in $\beta$ are shown by the two dotted lines $\langle \beta \rangle = 0.0$ (lower) or $\langle \beta \rangle = 0.4$ (upper).}
    \label{fig:M_v_M}
    \end{figure*}

    In Figure \ref{fig:M_v_M} we compare the 3D escape velocity masses with halo masses within $r_{200}$ ($M_{200}$). In the top left panel, we use the exact densities and potentials as measured using the particles (e.g., Equation \ref{eq:F_r}). These caustic mass estimates are nearly unbiased with a scatter of $13\%$. In the top right panel of Figure \ref{fig:M_v_M} we utilize equation \ref{eq:m_caustic_nfw}, where only the density profile is used to fit the 3D NFW concentrations and their errors. As expected from Figure  \ref{fig:r_0_compare} the NFW-inferred 3D caustic masses are biased low by $\sim$ 10\%. The scatter is 8\%. The errors on this panel use a conservative uncertainty in $c =$ 50\%. A large uncertainty in the concentration has little effect on the scatter of the actual caustic mass estimate. This will become important when we discuss realistic observational biases and scatters in section 3.3.

\subsection{Virialization}
\label{sec:test_virialization}
    It has been shown that the virial relation $2T = -W$ is often not met in simulated halos \citep{Shaw06,Bett07,Neto07,Davis11}. This does not mean that the system is not virialized, but simply that more terms from the tensor virial equation are required, usually in a surface pressure kinetic term. So the question then is at what radius to we begin to see a bias expected from equation \ref{eq:vesc_sigma} when $\langle b \rangle \ne 0$?
        
    To test this, we measure the exact (unbiased) caustic masses using $\mathcal{F} (r)$ at 1, 0.9, 0.7, and 0.5 $\times$ the virial radius. We detect no appreciable bias until we reach half a virial radius where the masses become biased low by 10\%.  We calculate $\langle b \rangle = 0.1$ for particles within this radius. We then apply $\langle b \rangle = 0.1$ during the virial calibration stage of the caustic technique and find no mass bias. \citet{Serra11} conduct a similar test, but against various fractions of their membership radius R$_{Tree}$, as opposed to an intrinsic cluster property like R$_{200}$. They find that there is a preferred radius of of 0.7$\times$R$_{Tree}$. We come to a slightly different conclusion: that the choice of radius does not matter, so long as the correct  $\langle b \rangle$ is used. We also find that there is no bias when caustic masses are calibrated using data within $0.7 \le R  \le 1 r_{200}$.
        
\subsection{Velocity Anisotropy}\label{sec:Projection}

    When the data are projected along the line-of-sight, velocity anisotropies in the orbits of the galaxies must be taken into account \citep{Diaferio97,Gifford13a}.  In the bottom left panel of  Figure \ref{fig:M_v_M}, we use $\mathcal{F}_{\beta}(r)$ which is the exact $\mathcal{F}(r)$ profile multiplied by the exact $\beta (r)$ profile. The increase in the scatter from the 3D ($\sim$ 10\%) case to the line-of-sight ($\sim$ 25\%) case is identical to what was measured in \citetalias{Gifford13a}, who use the classical caustic technique and a constant $\mathcal{F}_{\beta}$. Therefore, for any given cluster, there is no gain in accuracy or precision in the estimated caustic masses by measuring a $\beta (r)$ profile for each cluster. The scatter is dominated by line-of-sight variations in the projected velocity dispersion (see also \citetalias{Gifford13a}).
 
    We show our most realistic comparison of the caustic masses to $M_{200}$ in the bottom-right panel of Figure \ref{fig:M_v_M}. Here we drop explicit knowledge of the concentrations and apply the ensemble average $\langle c \rangle =5 \pm{2}$ for every halo in $\hat{\mathcal{F}}(r)$ (see Figure \ref{fig:r_0_compare}). We also drop explicit knowledge of the anisotropy profile and instead use $\langle \beta \rangle = 0.2 \pm{0.2}$ which is the average $\beta$ for these halos. Using these estimates, we find that the scatter is only slightly higher than in Figure \ref{fig:M_v_M} (bottom left). Large uncertainties in the average anisotropy and concentrations do not appreciably add scatter to what is already there from the line-of-sight projection. The bias in Figure \ref{fig:M_v_M} bottom-right is caused by the faulty assumption that the halos have the same NFW density and potential concentrations (see Figure \ref{fig:r_0_compare}).

    Systematic errors in the observable ensemble averages for the concentrations and the velocity anisotropies do impart mass biases. When we impose $\langle \beta \rangle = 0.0 \pm{0.2}$  the average bias changes from -10\% to -18\% while $\langle \beta \rangle = 0.4 \pm{0.2}$ results in an average mass bias of +5\%. When we impose $\langle c \rangle = 3 \pm{0.1}$  the bias changes from -8\% to -13\% while $\langle c \rangle = 6 \pm{2}$ results in a mass bias of -7\%. These average concentration values cover the full range of observational estimates in the literature \citep{Carlberg97,LinMohr04,Wojtak10,Budzynski12}.
      
\section{Conclusions}
\label{sec:Discussion}

    One goal for this {\it Letter} was to test the fundamental statistical and systematic precision of the escape velocity (or caustic) technique to measure masses of cluster-sized halos in N-body simulations. Given the 3D data, caustic masses are unbiased with 10-15\% precision (similar or better to the virial scaling relation of \citet{Evrard08}. The scatter increases to 25\% as a result of line-of-sight projections.
     
    Our second goal was to re-frame the theory in terms of observable quantities and remove calibrations to N-body simulations. We utilized the weak assumption that the observed density and potential profiles can be described by an NFW with the same shape parameters, specifically the scale parameter $r_0$. We find that this latter assumption does not hold in the Millennium Simulation data, and the inferred cluster masses are biased low by $\sim$ 10\%. The virial calibration can also contribute to biases in the caustic masses when the velocity dispersion is averaged over a radius where the total binding energy is not represented by virial expectations. We show that large uncertainties in the ensemble average of the velocity anisotropies and concentrations do not contribute significantly to the intrinsic line-of-sight scatter in projected caustic masses.  However, large  (e.g. 100\%) systematic errors in the average velocity anisotropies and concentrations can lead to additional 5-10\% biases in the caustic masses.

\section*{Acknowledgements}
The authors made use of the FLUX High Performance Computing Cluster at the University of Michigan.
The Millennium Simulation databases used in this paper and the web application providing online access to them were constructed as part of the activities of the German Astrophysical Virtual Observatory (GAVO). The authors want to especially thank Gerard Lemson for his assistance and access to the particle data and the referee for helpful comments. This material is based upon work supported by the National Science Foundation Graduate Student Research Fellowship under Grant No. DGE 1256260.

\bibliographystyle{apj}
%\bibliography{gifftex}{}

\begin{thebibliography}{27}
\expandafter\ifx\csname natexlab\endcsname\relax\def\natexlab#1{#1}\fi
\bibitem[{{Andreon} \& {Hurn}(2010)}]{Andreon10}
{Andreon}, S., \& {Hurn}, M.~A. 2010, \mnras, 404, 1922

\bibitem[{{Bett} {et~al.}(2007){Bett}, {Eke}, {Frenk}, {Jenkins}, {Helly}, \&
  {Navarro}}]{Bett07}
{Bett}, P., {Eke}, V., {Frenk}, C.~S., {Jenkins}, A., {Helly}, J., \&
  {Navarro}, J. 2007, \mnras, 376, 215

\bibitem[{{Binney} \& {Tremaine}(1987)}]{Binney87}
{Binney}, J., \& {Tremaine}, S. 1987, {Galactic dynamics} (Princeton, NJ: Princeton Univ. Press)

\bibitem[{{Biviano} \& {Poggianti}(2009)}]{Biviano09}
{Biviano}, A., \& {Poggianti}, B.~M. 2009, \aap, 501, 419

\bibitem[{{Budzynski} {et~al.}(2012){Budzynski}, {Koposov}, {McCarthy},
  {McGee}, \& {Belokurov}}]{Budzynski12}
{Budzynski}, J.~M., {Koposov}, S.~E., {McCarthy}, I.~G., {McGee}, S.~L., \&
  {Belokurov}, V. 2012, \mnras, 423, 104

\bibitem[{{Carlberg} {et~al.}(1997){Carlberg}, {Yee}, {Ellingson}, {Morris},
  {Abraham}, {Gravel}, {Pritchet}, {Smecker-Hane}, {Hartwick}, {Hesser},
  {Hutchings}, \& {Oke}}]{Carlberg97}
{Carlberg}, R.~G., {et~al.} 1997, \apjl, 485, L13

\bibitem[{{Chandrasekhar}(1969)}]{Chandra69}
{Chandrasekhar}, S. 1969, {Ellipsoidal figures of equilibrium} (New Haven, CT: Yale Univ. Press)

\bibitem[{{Conway}(2000)}]{Conway00}
{Conway}, J.~T. 2000, \mnras, 316, 555

\bibitem[{{Davis} {et~al.}(2011){Davis}, {D'Aloisio}, \& {Natarajan}}]{Davis11}
{Davis}, A.~J., {D'Aloisio}, A., \& {Natarajan}, P. 2011, \mnras, 416, 242

\bibitem[{{Diaferio}(1999)}]{Diaferio99}
{Diaferio}, A. 1999, \mnras, 309, 610

\bibitem[{{Diaferio} \& {Geller}(1997)}]{Diaferio97}
{Diaferio}, A., \& {Geller}, M.~J. 1997, \apj, 481, 633

\bibitem[{{Dressler} {et~al.}(2011){Dressler}, {Bigelow}, {Hare}, {Sutin},
  {Thompson}, {Burley}, {Epps}, {Oemler}, {Bagish}, {Birk}, {Clardy},
  {Gunnels}, {Kelson}, {Shectman}, \& {Osip}}]{Dressler11}
{Dressler}, A., {et~al.} 2011, \pasp, 123, 288

\bibitem[{{Evrard} {et~al.}(2008){Evrard}, {Bialek}, {Busha}, {White}, {Habib},
  {Heitmann}, {Warren}, {Rasia}, {Tormen}, {Moscardini}, {Power}, {Jenkins},
  {Gao}, {Frenk}, {Springel}, {White}, \& {Diemand}}]{Evrard08}
{Evrard}, A.~E., {et~al.} 2008, \apj, 672, 122

\bibitem[{{Geller} {et~al.}(2013){Geller}, {Diaferio}, {Rines}, \&
  {Serra}}]{Geller13}
{Geller}, M.~J., {Diaferio}, A., {Rines}, K.~J., \& {Serra}, A.~L. 2013, \apj,
  764, 58

\bibitem[{{Gifford} {et~al.}(2013){Gifford}, {Miller}, \& {Kern}}]{Gifford13a}
{Gifford}, D., {Miller}, C.~J., \& {Kern}, N. 2013, \apj

\bibitem[{{Guo} {et~al.}(2011){Guo}, {White}, {Boylan-Kolchin}, {De Lucia},
  {Kauffmann}, {Lemson}, {Li}, {Springel}, \& {Weinmann}}]{Guo11}
{Guo}, Q., {et~al.} 2011, \mnras, 413, 101

\bibitem[{{Host} {et~al.}(2009){Host}, {Hansen}, {Piffaretti}, {Morandi},
  {Ettori}, {Kay}, \& {Valdarnini}}]{Host09}
{Host}, O., {Hansen}, S.~H., {Piffaretti}, R., {Morandi}, A., {Ettori}, S.,
  {Kay}, S.~T., \& {Valdarnini}, R. 2009, \apj, 690, 358

\bibitem[{{Le F{\`e}vre} {et~al.}(2003){Le F{\`e}vre}, {Saisse}, {Mancini},
  {Brau-Nogue}, {Caputi}, {Castinel}, {D'Odorico}, {Garilli}, {Kissler-Patig},
  {Lucuix}, {Mancini}, {Pauget}, {Sciarretta}, {Scodeggio}, {Tresse}, \&
  {Vettolani}}]{LeFevre03}
{Le F{\`e}vre}, O., {et~al.} 2003, in Society of Photo-Optical Instrumentation
  Engineers (SPIE) Conference Series, Vol. 4841, Society of Photo-Optical
  Instrumentation Engineers (SPIE) Conference Series, ed. M.~{Iye} \& A.~F.~M.
  {Moorwood}, 1670--1681

\bibitem[{{Lemze} {et~al.}(2009){Lemze}, {Broadhurst}, {Rephaeli}, {Barkana},
  \& {Umetsu}}]{Lemze09}
{Lemze}, D., {Broadhurst}, T., {Rephaeli}, Y., {Barkana}, R., \& {Umetsu}, K.
  2009, \apj, 701, 1336

\bibitem[{{Lin} {et~al.}(2004){Lin}, {Mohr}, \& {Stanford}}]{LinMohr04}
{Lin}, Y.-T., {Mohr}, J.~J., \& {Stanford}, S.~A. 2004, \apj, 610, 745

\bibitem[{{Navarro} {et~al.}(1997){Navarro}, {Frenk}, \& {White}}]{NFW97}
{Navarro}, J.~F., {Frenk}, C.~S., \& {White}, S.~D.~M. 1997, \apj, 490, 493

\bibitem[{{Neto} {et~al.}(2007){Neto}, {Gao}, {Bett}, {Cole}, {Navarro},
  {Frenk}, {White}, {Springel}, \& {Jenkins}}]{Neto07}
{Neto}, A.~F., {et~al.} 2007, \mnras, 381, 1450

\bibitem[{{Postman} {et~al.}(2012){Postman}, {Coe}, {Ben{\'{\i}}tez},
  {Bradley}, {Broadhurst}, {Donahue}, {Ford}, {Graur}, {Graves}, {Jouvel},
  {Koekemoer}, {Lemze}, {Medezinski}, {Molino}, {Moustakas}, {Ogaz}, {Riess},
  {Rodney}, {Rosati}, {Umetsu}, {Zheng}, {Zitrin}, {Bartelmann}, {Bouwens},
  {Czakon}, {Golwala}, {Host}, {Infante}, {Jha}, {Jimenez-Teja}, {Kelson},
  {Lahav}, {Lazkoz}, {Maoz}, {McCully}, {Melchior}, {Meneghetti}, {Merten},
  {Moustakas}, {Nonino}, {Patel}, {Reg{\"o}s}, {Sayers}, {Seitz}, \& {Van der
  Wel}}]{Postman12}
{Postman}, M., {et~al.} 2012, \apjs, 199, 25

\bibitem[{{Rines} {et~al.}(2013){Rines}, {Geller}, {Diaferio}, \&
  {Kurtz}}]{Rines12}
{Rines}, K., {Geller}, M.~J., {Diaferio}, A., \& {Kurtz}, M.~J. 2013, \apj,
  767, 15

\bibitem[{{Serra} {et~al.}(2011){Serra}, {Diaferio}, {Murante}, \&
  {Borgani}}]{Serra11}
{Serra}, A.~L., {Diaferio}, A., {Murante}, G., \& {Borgani}, S. 2011, \mnras,
  412, 800

\bibitem[{{Shaw} {et~al.}(2006){Shaw}, {Weller}, {Ostriker}, \&
  {Bode}}]{Shaw06}
{Shaw}, L.~D., {Weller}, J., {Ostriker}, J.~P., \& {Bode}, P. 2006, \apj, 646,
  815

\bibitem[{{Springel} {et~al.}(2005){Springel}, {White}, {Jenkins}, {Frenk},
  {Yoshida}, {Gao}, {Navarro}, {Thacker}, {Croton}, {Helly}, {Peacock}, {Cole},
  {Thomas}, {Couchman}, {Evrard}, {Colberg}, \& {Pearce}}]{Springel05}
{Springel}, V., {et~al.} 2005, \nat, 435, 629

\bibitem[{{Wojtak} \& {{\L}okas}(2010)}]{Wojtak10}
{Wojtak}, R., \& {{\L}okas}, E.~L. 2010, \mnras, 408, 2442
\end{thebibliography}

\end{document}